\documentclass[]{spie}  

\usepackage{amsmath,amsfonts,amssymb}
\usepackage{graphicx}
\usepackage[table,xcdraw]{xcolor}
\usepackage[colorlinks=true, allcolors=blue]{hyperref}
\usepackage[export]{adjustbox}
\usepackage{caption}
\usepackage{subcaption}
\usepackage[numbers]{natbib}

\title{X-ray speed reading: enabling fast, low noise readout for next-generation CCDs}

\author[a]{S. Herrmann}
\author[a]{P. Orel}
\author[a]{T. Chattopadhyay}
\author[a]{R. G. Morris}
\author[b]{G. Prigozhin}
\author[c]{K. Donlon}
\author[b]{R. Foster}
\author[b]{M. Bautz}
\author[a,d,e]{S. Allen}
\author[c]{C. Leitz}

\affil[a]{Kavli Institute of Astrophysics and Cosmology, Stanford University, 452 Lomita Mall, Stanford, CA 94305, USA}
\affil[b]{Kavli Institute for Astrophysics \& Space Research, Massachusetts Institute of Technology, \hspace{4em} 77 Massachusetts Avenue, Cambridge, MA 02139, USA}
\affil[c]{MIT Lincoln Laboratory, 244 Wood Street, Lexington, MA 20420, USA}
\affil[d]{Department of Physics, Stanford University, 382 Via Pueblo Mall, Stanford CA 94305, USA}
\affil[e]{SLAC National Accelerator Laboratory, 2575 Sand Hill Road, Menlo Park, CA 94025, USA}

\authorinfo{Send correspondence to svench@stanford.edu}

\begin{document} 
\maketitle

\begin{abstract}
Current, state-of-the-art CCDs are close to being able to deliver all key performance figures for future strategic X-ray missions except for the required frame rates. Our Stanford group is seeking to close this technology gap through a multi-pronged approach of microelectronics, signal processing and novel detector devices, developed in collaboration with the Massachusetts Institute of Technology (MIT) and MIT Lincoln Laboratory (MIT-LL). Here we report results from our (integrated) readout electronics development, digital signal processing and novel SiSeRO (Single electron Sensitive Read Out) device characterization.
\end{abstract}


\section{Introduction}
\label{sec:intro}  
\begin{figure}[ht]
\begin{subfigure}{.28\textwidth}
  \centering
  \includegraphics[width=.8\linewidth]{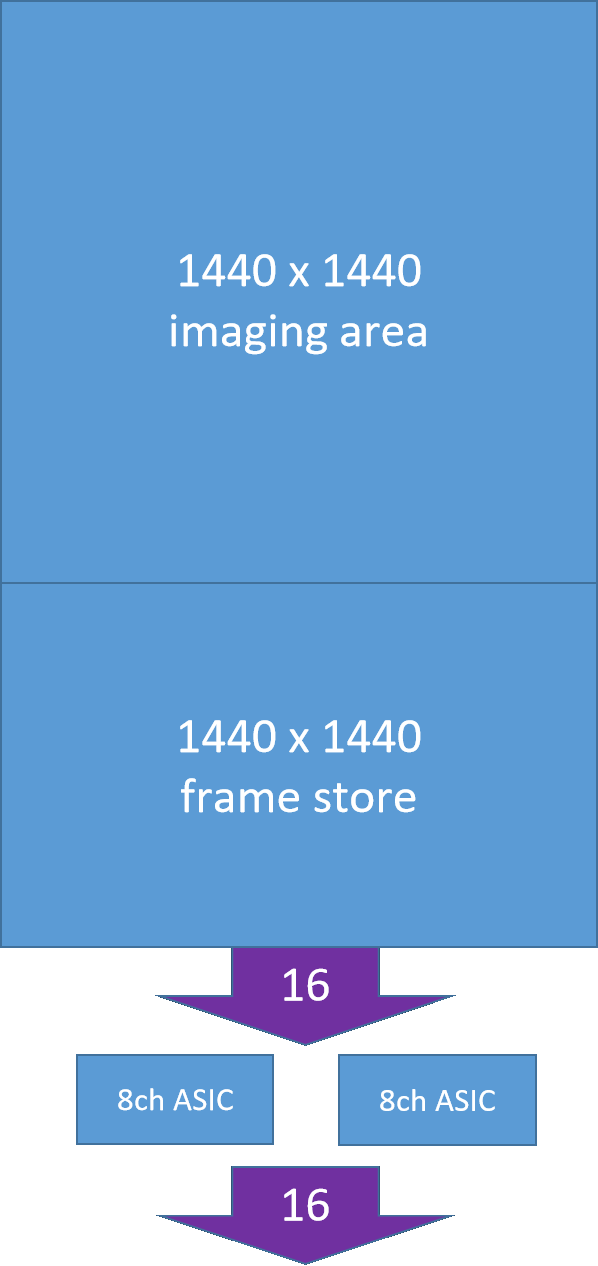}  
  \caption{Concept of a CCD module}
  \label{fig:AXISCCD}
\end{subfigure}
\begin{subfigure}{.7\textwidth}
    \centering
    \begin{tabular}{|
    >{\columncolor[HTML]{FFFFFF}}l |c|c|}
    \hline
    \multicolumn{1}{|c|}{\cellcolor[HTML]{2277B8}{\color[HTML]{FFFFFF} \textbf{Parameter}}} & 
    \cellcolor[HTML]{2277B8}{\color[HTML]{FFFFFF}  \textbf{ \vtop{\hbox{\strut demonstrated}\hbox{\strut lab capability}} } }& 
    \cellcolor[HTML]{2277B8}{\color[HTML]{FFFFFF} \textbf{target capability}} \\ 
    \hline
    CCD size [pixels] & 1440 & 1440 \\ 
    \hline
    pixel size [um] & 24 & 24 \\ 
    \hline
    total FOV of 4 modules [armin '] & 25.1 & 25.1 \\ 
    \hline
    CCD outputs & 8 & 16 \\
    \hline
    IA to DS clock rate [MHz] & 1 & 2 \\
    \hline
    serial pixels per output  & 160 & 90 \\
    \hline
    pixel rate per output [MPix/s] & 2 & 4  \\ 
    \hline
    FS stransfer time [ms] & 1.4 & 0.7  \\ 
    \hline
    readout time [ms] & 140 & 37  \\ 
    \hline
    frame rate [fps] & 7 & 26  \\ 
    \hline
    out of time ratio & 1.0\%  & 1.9\%  \\ 
    \hline
    photon rate per sec and pixel & 2.4 & 8.7 \\ \hline
    
    \end{tabular}
    
  \caption{specifications and estimated performance }
  \label{fig:AXIStable}
\end{subfigure}
\caption{One possible implementation for a AXIS focal plane would be to tile four sub modules each consisting of a frame-transfer CCD with 1440 x 1440 pixels of 24 $\mu m$ size into a detector with a total of 8 Mega-pixels, where each pixel corresponds to 0.5”. 
Image to frame store transfer could be achieved in roughly 1ms, with read out of the frame store occurring while the subsequent image is gathered. To achieve a desired frame rate of more than 20 fps, the CCD would use 16 outputs distributed across the 34mm of the detector edge, each of them read out with a pixel rate of at least 4 MHz.}
\label{fig:AXISmodule}
\end{figure}

Future strategic X-ray astronomy missions like AXIS \cite{mushotzky2019_axis} propose a combination of large collecting area mirrors with large, fast, wide-field imagers. High frame rates will be essential to minimize the impact of pile-up for point sources and to mitigate the impact of the particle background on studies of faint, diffuse gas. At the same time, low noise and excellent soft X-ray energy response must also be maintained to meet the key science goals. State-of-the-art CCDs are close to being able to deliver all key performance figures for such missions except for the required frame rates. Fast frame rates for large detectors result in very high effective pixel rates. Our group at Stanford is addressing this technology gap through a multi-pronged approach, in collaboration with the Massachusetts Institute of Technology (MIT) and MIT Lincoln Laboratory (MIT-LL). To achieve higher frame rates, we are working to increase both the readout speed of individual outputs and the number of outputs per CCD that can operate in parallel. Figure \ref{fig:AXISmodule} shows a possible CCD module concept suitable for the AXIS focal plane. Speed increases on individual outputs stem from CCD output stage optimization, reduction of parasitic output loading through the use of an dedicated ASIC, and the use of digital signal processing on the video waveform. The readout ASIC also allows us to operate multiple outputs in parallel with a small footprint and modest power consumption. We are also investigating a novel detector technology manufactured at MIT-LL, the Single electron Sensitive Read Out (hereafter SiSeRO) that, while not quite yet capable of single electron noise performance, offers a promising pathway to very low noise, high speed X-ray detectors.

\section{CCD outputs and video signal}

\begin{figure}[ht]
\begin{subfigure}{.5\textwidth}
  \centering
  \includegraphics[width=0.95\linewidth]{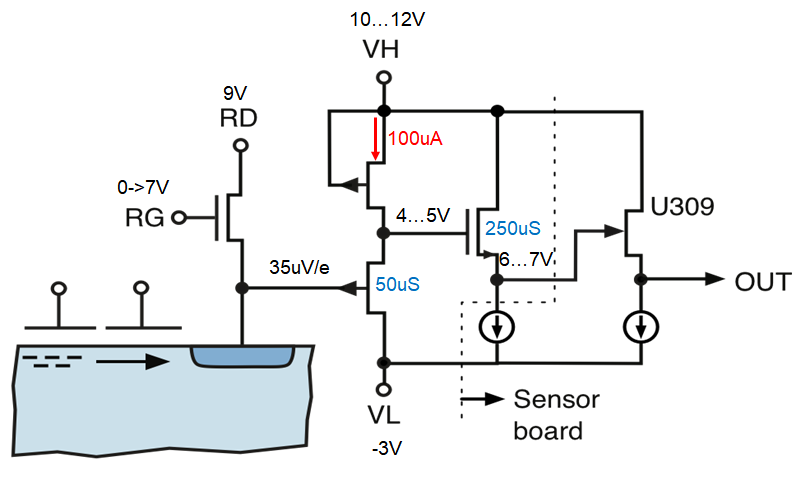}  
  \caption{CCD output stage schematic}
  \label{fig:CCDschem}
\end{subfigure}
\begin{subfigure}{.5\textwidth}
  \centering
  \includegraphics[width=1.0\linewidth]{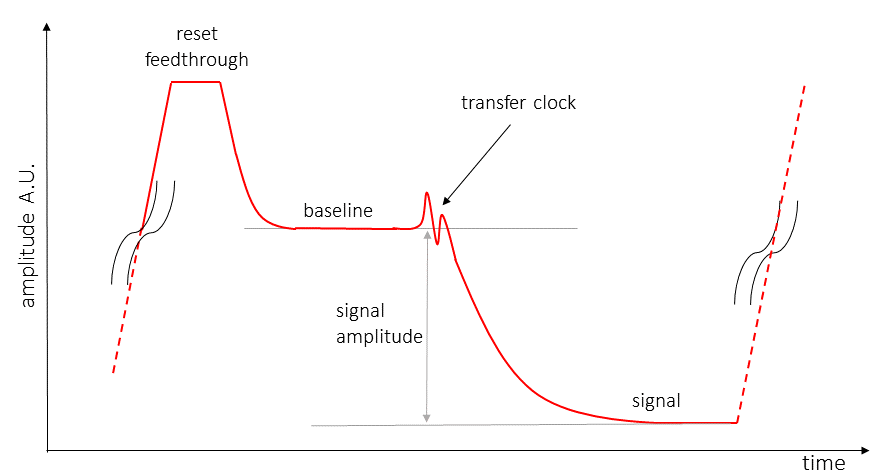}
  \caption{typical video waveform}
  \label{fig:CCDtypWF}
\end{subfigure}
\caption{
Left - CCD output stage schematic consisting of a floating diffusion, a JFET first stage followed by a NMOS second stage and a discrete transistor third stage.
Right - A typical CCD video waveform seen at the source of the output stage transistor. The output exhibits usually a large DC offset from the bias points of the output stage. During the readout operation the waveform goes through the stages of reset, baseline settling, charge transfer and signal settling for every pixel that is read out}
\label{fig:CCDtyp}
\end{figure}

The output of a CCD is usually implemented with a floating diffusion, which collects the charge signal, connected to the gate of a source follower biased with a current source. While the gate is not driven by a low impedance, as for a normal source follower, but rather by a high impedance floating capacitor plate, any charge moved onto this node will change the output voltage at the source. In order to achieve a large conversion gain from charge to voltage, the node capacitance needs to be minimized, which demands a small transistor with minimal gate capacitance and therefore limited transconductance. To improve the drive strength (and speed) of such an output, further source follower stages utilizing larger transistors are often used. To establish a bias point for the floating capacitor plate and therefore the source of the transistor, the node is periodically reset to a fixed potential - this potential is commonly called the Reset-Drain voltage, supplied via a reset transistor (switch). Figure \ref{fig:CCDschem} shows such an output for a MIT Lincoln Laboratory CCD with a p-JFET first stage and a NMOS second stage. The first stage is biased with a CCD on-chip current source, while the second stage requires an off-chip current source.  During operation, charge is shifted through the CCD registers and will end on the floating diffusion and modulate the transistor gate. The whole process is depicted in Figure \ref{fig:CCDtypWF} and consists of the following steps, with the resulting waveform at the output commonly called the CCD video signal:
\begin{itemize}
\setlength{\itemsep}{-2pt}
\item Reset - the reset pre-charges the floating diffusion, and the gate of the first stage, to a predefined potential. The reset clock on the gate of the reset transistor often generates considerable reset feed-through on the video waveform.
\item Baseline settling - after the reset, the output will settle to a baseline voltage level that corresponds to the empty charge state.
\item Charge transfer – clocking transfers the charge from the serial register of the CCD to the floating diffusion and, therefore, the gate of the first stage. This clocking often generates a disturbance on the video waveform due to parasitic capacitive coupling.
\item Signal settling - after the signal charge is transferred and acts on the transistor gate, the output will settle to a signal level. The change from the baseline to the signal level is directly proportional to the amount of charge transferred. If the charge packet consists of electrons, the resulting signal level is at a lower potential than the baseline level.
\end{itemize}

The signal amplitude and, therefore, the size of the charge packet transferred is extracted from the waveform by taking the difference between the signal and baseline, more recently performed by  digital signal processing of the sampled waveform.
The described voltage readout with source follower output stage is a common implementation, but not the only option. Another option is drain current readout, as pioneered by the Athena Wide Field Imager\cite{10.1117/12.2054490} and its readout chip VERITAS\cite{10.1117/12.2056097}. In the MIT Lincoln Laboratory SiSeRo (Single-electron Sensitive Readout) implementation, the floating diffusion and the JFET are replaced with a PMOS transistor that includes a region under the transistor (in the bulk of the silicon) that serves as the back gate. The CCD channel can clock the charge signal into the region under the PMOS and also move it out again. Any charge transferred in the area under the PMOS modulates the transistor channel changing the current flow.  To optimize charge transfer in and out of the transistor area, fixed voltages at the transistor terminals are preferred. At this fixed voltage a small current biases the transistor and the signal is determined by measuring the change in current.

\section{increasing the readout speed for pJFET outputs }

The output stage as outlined above has multiple nodes that limit the speed due to their inherent RC time constants. The first limitation is the pJFET driving the second stage NMOSFET. In the CCID85 there is roughly a five fold difference in Gm between the two transistors, slightly higher than the usual factor of three that is commonly used. (It is to be noted that these are also two different types of transistors and such the optimum scaling is different.) We estimate this limitation to be around 8 MHz in bandwidth. The next limitation is the second stage driving the off-chip stage. From DC characterization we estimate the $g_{m}$ of the NMOS to be around 250 $\mu S$ (depending on the exact bias condition). For the following measurements, we have been using our TinyBox CCD \cite{chattopadhyay20_spie} test system that utilizes an STA Archon Controller with digital correlated double sampling but also a raw mode to export complete video waveforms for offline processing. 

\subsection{Improved discrete electronics}
\begin{figure} [ht]
\centering
\includegraphics[width=0.8\textwidth,keepaspectratio=tru]{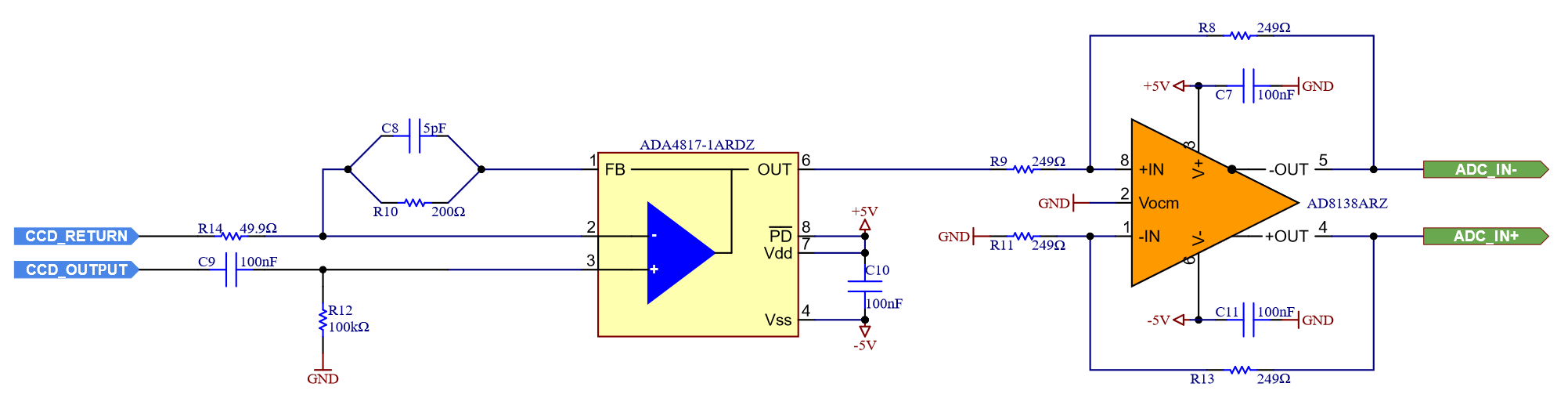}
\vspace*{0.1mm}
\caption[ADA4817PRE] 
{ \label{fig:ADA4817PRE} 
Schematic of the preamplifier circuit to read out the CCD analog output. It uses a single-ended operational amplifier (ADA4817) in a non-inverting configuration as the first stage, which is followed by a fully differential ADC driver (AD8138). }
\end{figure}

In order to minimize the load capacitance and not add further voltage follower stages that exhibit gains below unity and small but not negligible noise components, we decided to replace the external JFET with an active amplification stage with small input capacitance. 
We selected the ADA4817 OPAMP for its wide bandwidth (1 GHz), small input capacitance (1.3 pF) and low input noise (4 nV/$\surd$Hz). The complete circuit is shown in figure \ref{fig:ADA4817PRE} and delivers an effective bandwidth of 50MHz (including compensation) at a gain of 5, with a total input referred noise of around 15 nV/$\surd$Hz.

\begin{figure}[ht]
    \centering
    \subfloat[CCD waveforms]{
             \includegraphics[width=0.52\textwidth,keepaspectratio=tru]{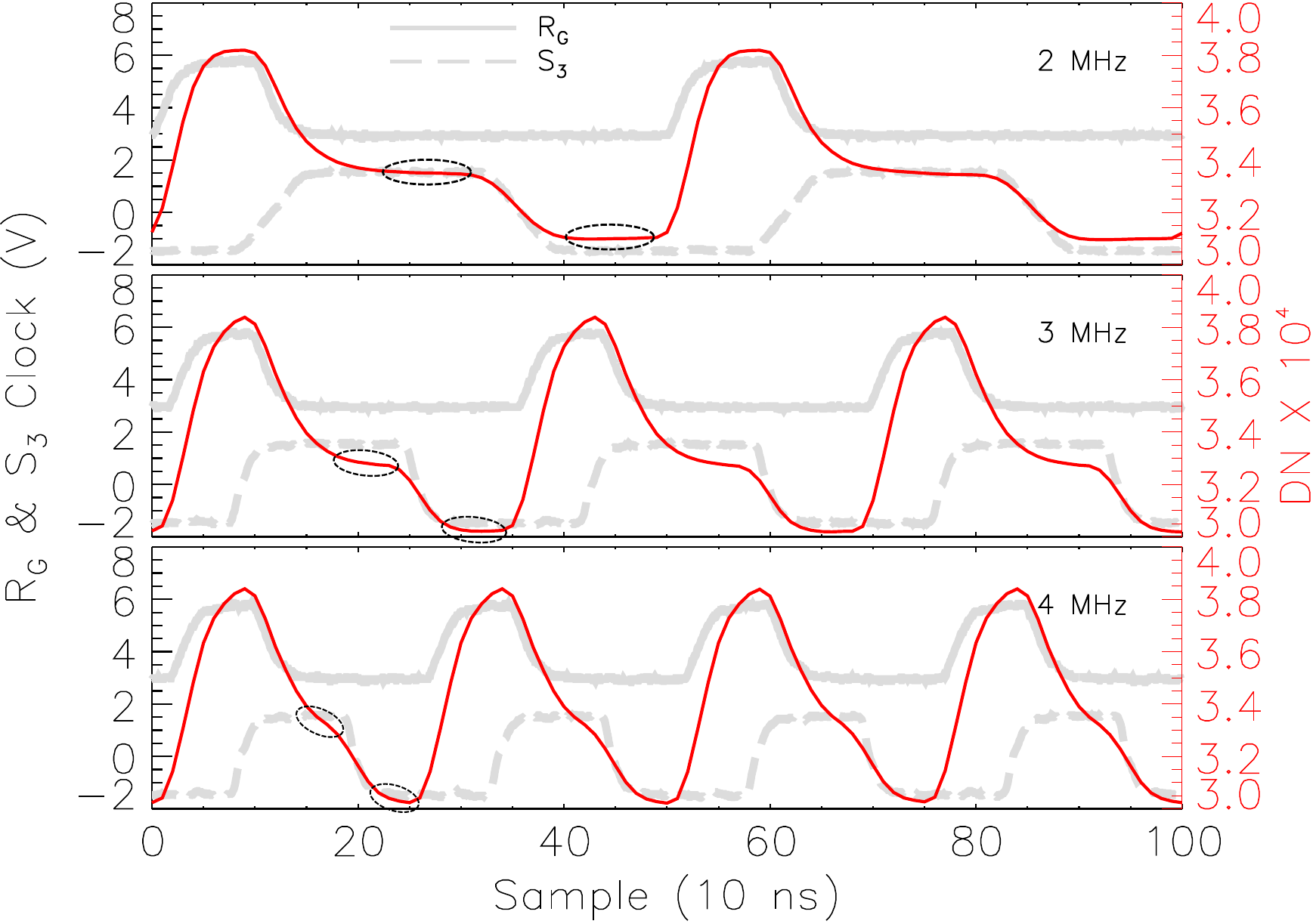}
    }
    \subfloat[measurement results]{
        \begin{tabular}{|l|l|l|l|} 
            \hline
            \rule[-1ex]{0pt}{3.0ex}  Readout speed (MHz) & 2 & 3 & 4  \\
            \hline\hline
            \rule[-1ex]{0pt}{3.0ex}  Num. of samples in CDS &18 &10 &7    \\
            \hline
            \rule[-1ex]{0pt}{3.0ex}  Read noise ($e^{-}_{RMS}$) &5.3&6.1&7.6  \\
            \hline
            \rule[-1ex]{0pt}{3.0ex}  Board noise ($e^{-}_{RMS}$) &1.7&2.0 & 3.2   \\
            \hline
            \rule[-1ex]{0pt}{3.0ex}  System gain ($ADU/e{^-}$) &0.84&0.83 &0.81    \\
            \hline
            \rule[-1ex]{0pt}{3.0ex}  Conversion gain ($\mu V/e^{-}$) &28&27.8&26.8    \\
            \hline
            \rule[-1ex]{0pt}{3.0ex}  FWHM (eV) @ 5.9 keV &133&139 &148    \\
            \hline
        \end{tabular}
        \newline
        \newline
        \newline
    }
     \caption{ \label{fig:CCID85results}
        Results from CCID85 characterisation at readout speeds of 2, 3 and 4 MPIX/sec. At high speeds the reset pulse dominates the readout time, resulting in the number of samples that can be used for signal processing dropping from 18 to 7, while the speed doubles from 2 to 4 MPix/sec. The readout noise increases by the expected factor of approximately $\sqrt{2}$ as the speed doubles.}
\end{figure}

While the input noise of the alternative JFET can be lower, one has to consider that a source follower stage has less then unity gain, requiring an additional gain stage in any case. In conclusion, the amplification stage performs as good as the solution with an external JFET but with smaller input capacitance, such that in a practical implementation the load capacitance will be limited by the packaging and PCB parasitic capacitance.
We performed an extensive characterisation campaign with a MIT Lincoln CCID85 in \cite{chattopadhyay22_ccd}, where we also optimized for high speed readout up to 4 MPix/sec, double that previously obtained. Figure \ref{fig:CCID85results} is a summary for the achieved results. As we increase the pixel rate, the video waveform gets squeezed and the baseline and signal region are barely recognizable. Further, the reset pulse begins to dominate the readout time, a problem we plan to improve with a faster reset clock driver in the next generation of the readout PCB. Due to the squeezed waveform, the available number of samples that can be used for digital CDS reduces from 18 to 7 as the speed doubles from 2 to 4 MPix/sec, resulting in a noise increase of roughly a factor of $\sqrt{2}$, in line with the expectations. The CCID85 used here has a conversion gain of 28($\mu V/e^{-}$) while the newer CCID93 delivers up to 60 ($\mu V/e^{-}$) and has demonstrated noise performance as low as 2.6 $e^{-}_{RMS}$ for 2 Mpix/sec \cite{Prigozhinetal2022}.

\subsection{MIT CCD Readout ASIC MCRC}

\begin{figure}[ht]
\begin{subfigure}{.5\textwidth}
  \centering
  \includegraphics[width=0.95\linewidth]{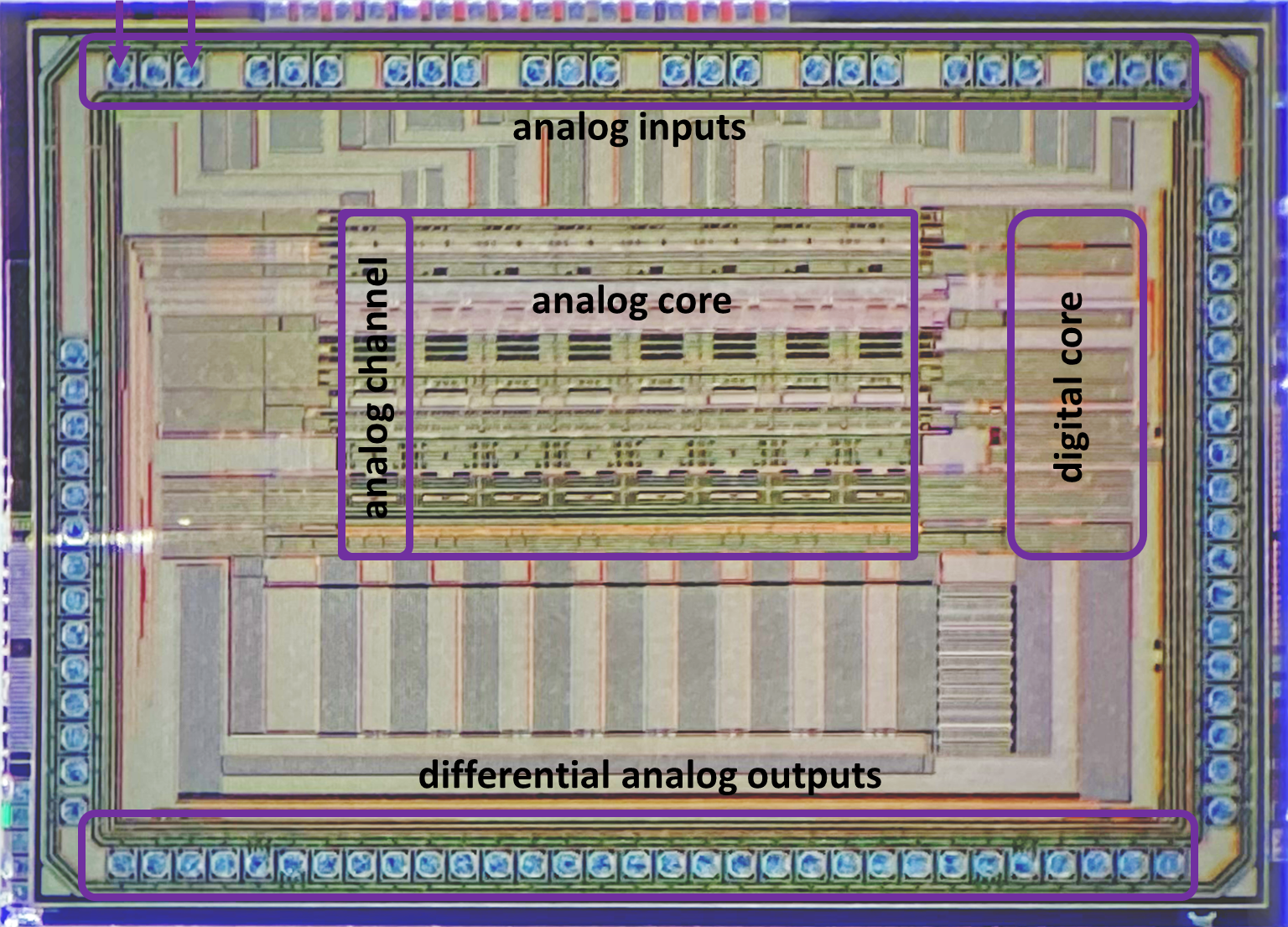}  
  \caption{Micro photograph of  the MCRC V1.0 chip}
  \label{fig:MCRCpic}
\end{subfigure}
\begin{subfigure}{.5\textwidth}
  \centering
  \includegraphics[width=1.0\linewidth]{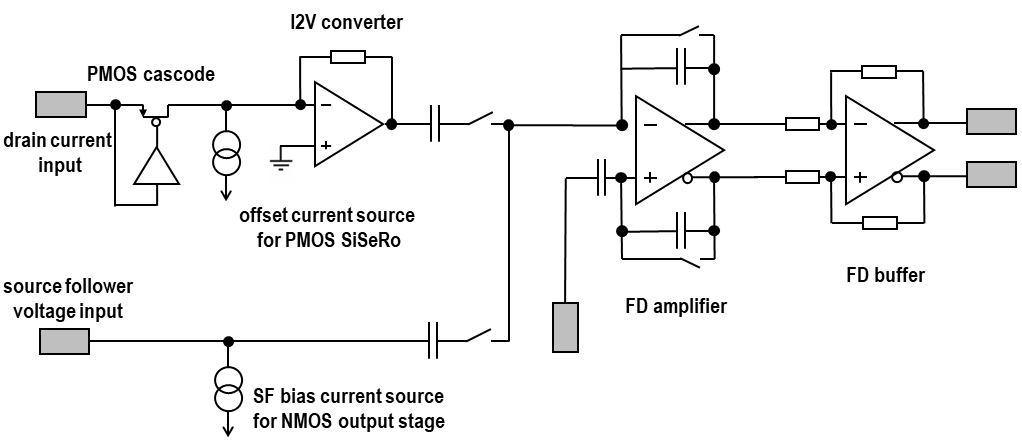}
  \caption{Overview schematic of the MCRC analog channel}
  \label{fig:MCRCanach}
\end{subfigure}
\caption{
Left - Photograph of the MCRC V1.0 chip. The eight groups of pads at the top are the analog inputs to the eight channels. The bottom row is composed of differential outputs pads and power supply pads, respectively. Pads on the right side belong to the digital communication interface, while the left side pads are part of the biasing and debugging circuits. The analog core is compact and located in the center of the chip.
Right - Overview schematic of the MCRC analog channel. Each of the analog channels consists of an input stage, which is user selectable between the voltage input and the current input. The input section of the chip is followed by a pre-amplifier and a fully differential buffer driver.}
\label{fig:MCRCoverview}
\end{figure}

Further improvements to the discrete readout electronics pose a number of challenges. Firstly, discrete electronics require considerable board space and power consumption. This is especially problematic when, in order to increase the frame-rate further, many parallel operating outputs are required. In addition, it is not obvious how the load capacitance to the detector output could be further reduced when packaging and PCB parasitics dominate the load capacitance. An integrated microelectronics solution can remedy all of these problems: 
\begin{itemize}
\setlength{\itemsep}{-2pt}
    \item The footprint is small enabling the ASIC to sit close to the detector output.
    \item The proximity of the unpackaged die allows the chip to be wire-bonded directly to the detector outputs, significantly reducing the load capacitance.
    \item Custom ASIC solutions usually have a much smaller power consumption than generic commercial components.
\end{itemize}

These potential advantages lead our group to developed the MIT CCD Readout Chip (MCRC) V1.0 \cite{herrmann20_mcrc}. Figure \ref{fig:MCRCpic} shows the manufactured chip and its functional blocks outlined on the photograph. The ASIC includes 8 readout analog channels operating in parallel, a digital SPI interface to program the ASIC settings and control the internal switch logic, four bias DACs to set the various internal circuitry operating points. Each of the analog channels consists of an input stage, which offers the user the ability to select between the voltage and the current inputs, respectively. In addition, both inputs feature CCD bias current sources. The input stage is followed by a pre-amplifier and a fully differential buffer driver. The chip can handle up to 12 V of input DC voltage at its voltage input. Figure \ref{fig:MCRCanach} shows an overview schematic of a single analog readout channel. The manufactured chip has been verified and characterised for its functionality and performance. Details of this characterisation can be found in \cite{Oreletal2022}. Compared to a discrete solution, the MCRC V1.0 benefits from:
\begin{itemize}
\setlength{\itemsep}{-2pt}
    \item A small input capacitance of only 1.5 pF total.
    \item A bandwidth of 50 MHz over the full chain.
    \item Low power consumption of only 35 mW per Channel. Roughly an order of magnitude less than the discrete implementation.
    \item A measured input noise voltage spectral density of 6.5 nV/$\sqrt{Hz}$. Further outperforming the discrete implementation.
\end{itemize}

\begin{figure} [ht]
\includegraphics[width=0.5\textwidth,center,keepaspectratio=tru]{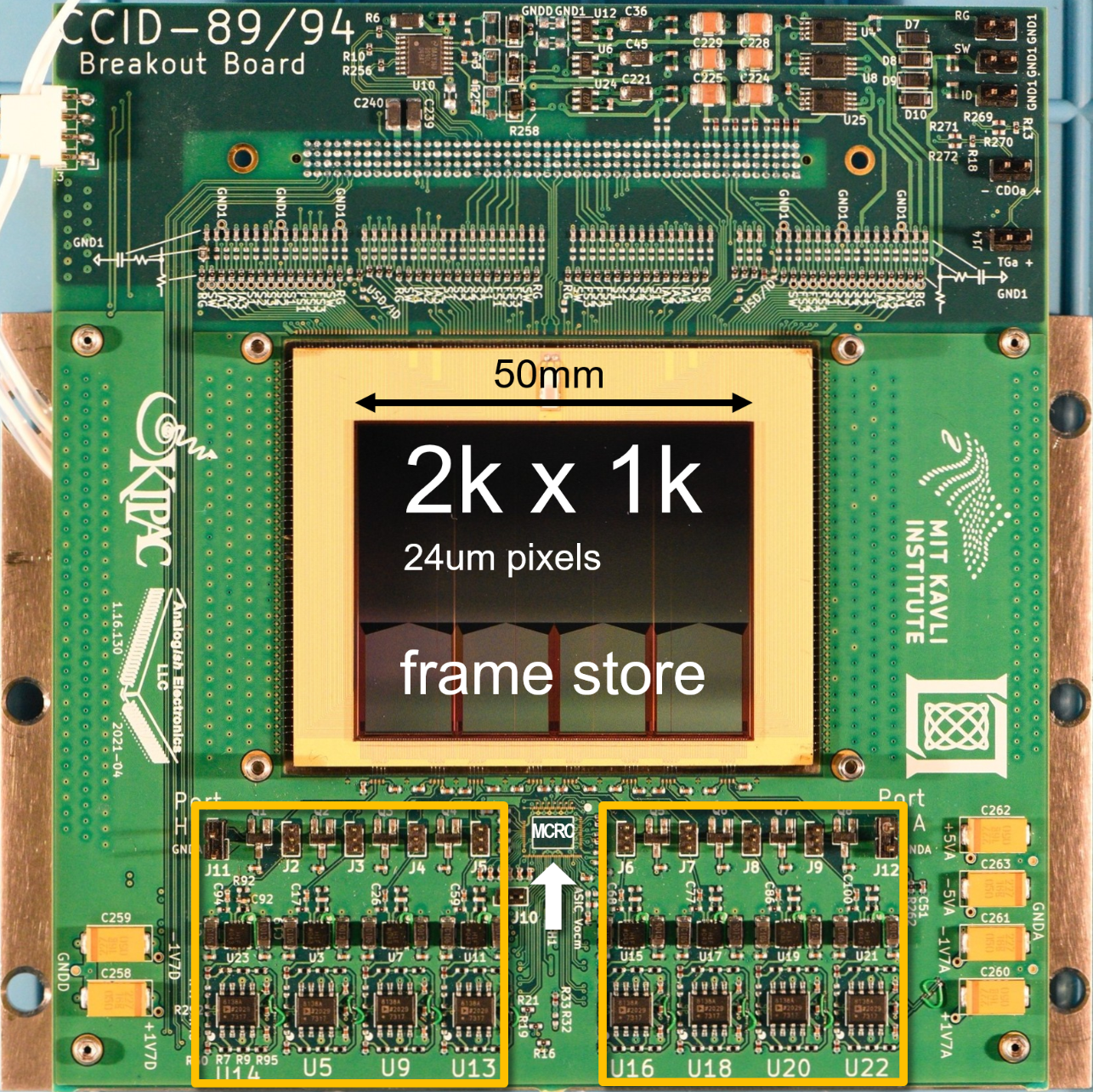}
\caption[CCID89module] 
{ \label{fig:CCID89module} 
Photography of a CCID89 X-ray CCD mounted in a  test circuit board. The large 2 megapixel sensor features eight output ports. The orange boxes show the 8-channel discrete readout electronics, while the MCRC V1.0 ASIC can be mounted  on the board mid-way along the bottom edge of the sensor. The replacement of the discrete electronics with the MCRC V1.0 ASIC will result in a substantial reduction in foot print and an order of magnitude power reduction.}
\end{figure}

The next stage of development is to test a multi-output CCD with the MCRC V1.0 ASIC. We have constructed a dedicated test module for the large 2 megapixel CCID89 sensor from MIT Lincoln Laboratory that features eight outputs. Figure \ref{fig:CCID89module} shows a photograph of the module. The MCRC V1.0 ASIC can be mounted on the board mid-way along the bottom edge of the sensor, replacing the eight-channel discrete electronics (highlighted in orange), enabling a substantial reduction in board space. Initial measurements with the discrete electronics have already started \cite{Bautzetal2022} and measurements with the ASIC will commence in the fall of this year. This module should provide frame rates of around 5 fps with the discrete electronics and 18 fps with the MCRC V1.0 ASIC.

\section{Digital Signal Processing}
Our readout concept offers the capability to examine the video waveform and perform digital signal processing, beyond the classical per pixel processing required to construct X-ray events and an image. 

\subsection{Digital signal processing on CCD with JFET output}

\begin{figure}[ht]
\begin{subfigure}{.32\textwidth}
  \centering
  \includegraphics[width=1.0\linewidth]{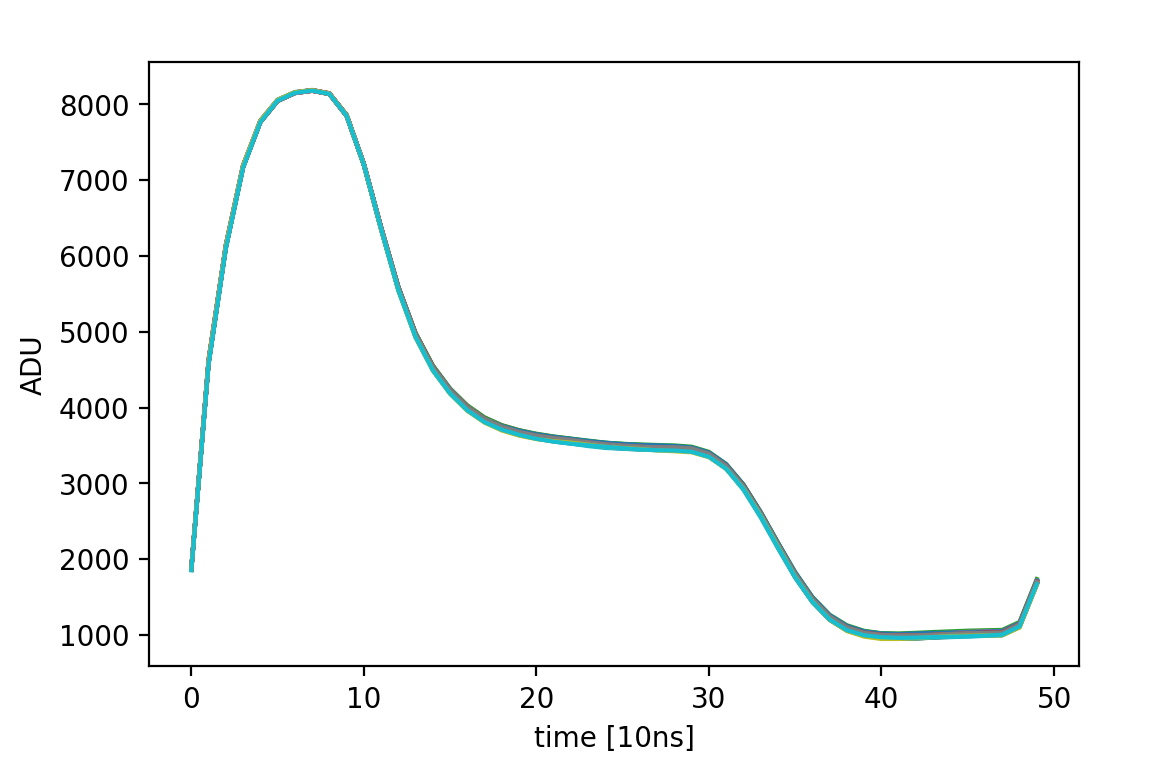}  
  \caption{ensemble of 50 waveforms (JFET)}
  \label{fig:JFET_WF_Raw}
\end{subfigure}
\begin{subfigure}{.32\textwidth}
  \centering
  \includegraphics[width=1.05\linewidth]{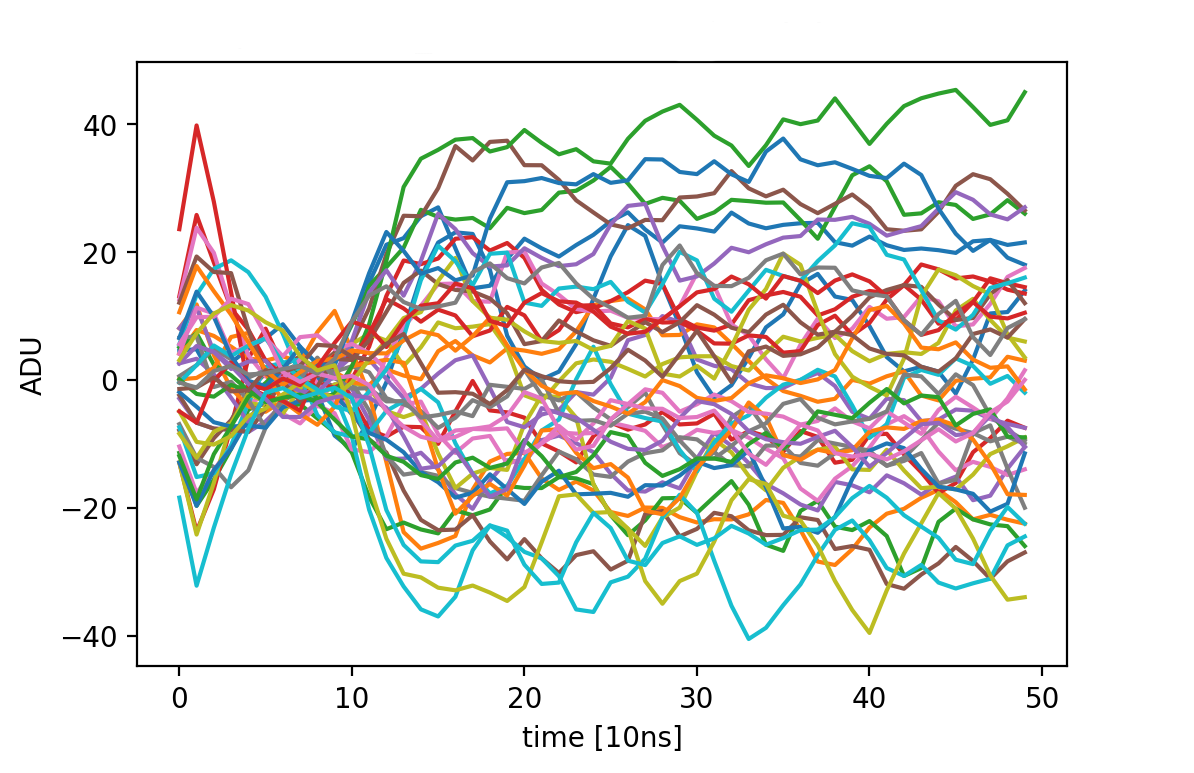}
  \caption{residuals of the 50 waveform}
  \label{fig:JFET_WF_Res}
\end{subfigure}
\begin{subfigure}{.37\textwidth}
  \centering
  \includegraphics[width=1.05\linewidth]{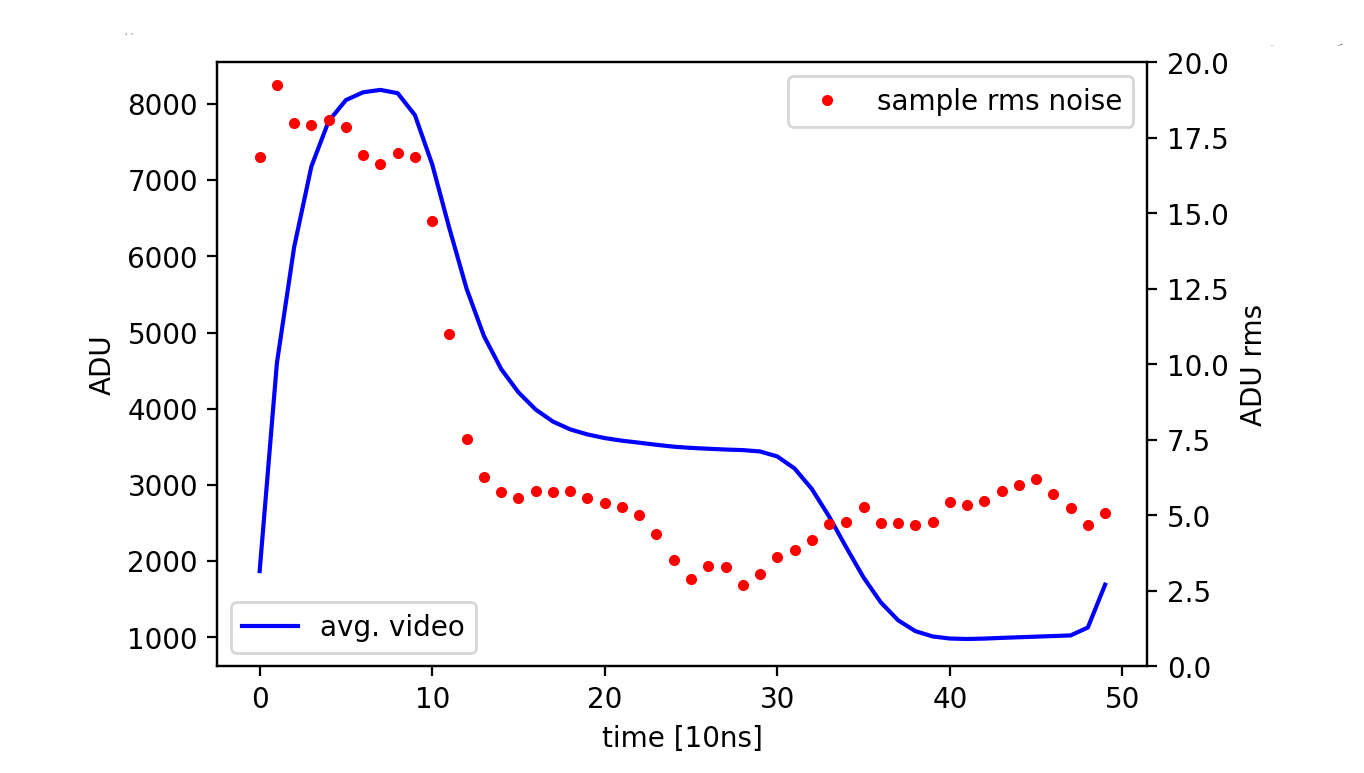}
  \caption{rms sample variation}
  \label{fig:JFET_WF_RMS}
\end{subfigure}
\caption{
Left - Video waveform ensemble of 50 pixels with the JFET output, in which the small variation between the waveforms can be barely seen.
Middle - Residuals of those 50 waveforms after subtraction of the average. During the reset pulse (at around 50-100 ns) the variation in the waveforms is small, while after the reset the variation is large due to KTC noise from the reset process (around 18 ADU rms). Right - root-mean-square variation of the samples across the 50 waveforms after each waveform has been clamped at the baseline level. The variation is low and approximately constant from baseline to signal, including during charge transfer, but raises sharply during the reset. }
\label{fig:JFET_WF}
\end{figure}

In order to optimize the digital CDS filtering of the 2 to 4 MPIX/sec measurements presented in section 3.1, we saved a number of raw video waveforms for offline processing. The first thing we examined was the overall variation of the waveforms from pixel to pixel, as most of this variation will be suppressed by the CDS filtering used for X-ray images.  Figure \ref{fig:JFET_WF_Raw} shows a waveform ensemble from 50 pixels, read out at a rate of 2 MPix/sec, which line up closely on top of each other. Figure \ref{fig:JFET_WF_Res} shows the residuals of the same data after the subtraction of the average waveform: The spread during the reset pulse (at around 50-100 ns) is small (around 3.5 ADU rms) while after the reset the output node settles at different levels (18 ADU rms), dominated by the KTC noise of the reset process. Due to the limited bandwidth of the output stage, the transient behavior between reset and baseline level is visible. The next question posed is how close the baseline measurement can be positioned towards the reset pulse. For this investigation, every single one of the 50 waveforms was offset corrected, such that the level at the baseline portion of the waveform is set to the average level. This process is equivalent to analog clamping. Then the rms variation for the waveform sample points across all the 50 waveforms was calculated. This will reflect the actual readout procedure, where the signal is measured relative to the baseline. Figure \ref{fig:JFET_WF_RMS} shows the resulting plot. The variation around the baseline portion is low, stays relatively constant between baseline and signal, but rises sharply as it approaches the reset pulse.

\subsection{Digital signal processing with SiSeRO outputs}

\begin{figure}[ht]
\begin{subfigure}{.45\textwidth}
  \centering
  \includegraphics[width=1.0\linewidth]{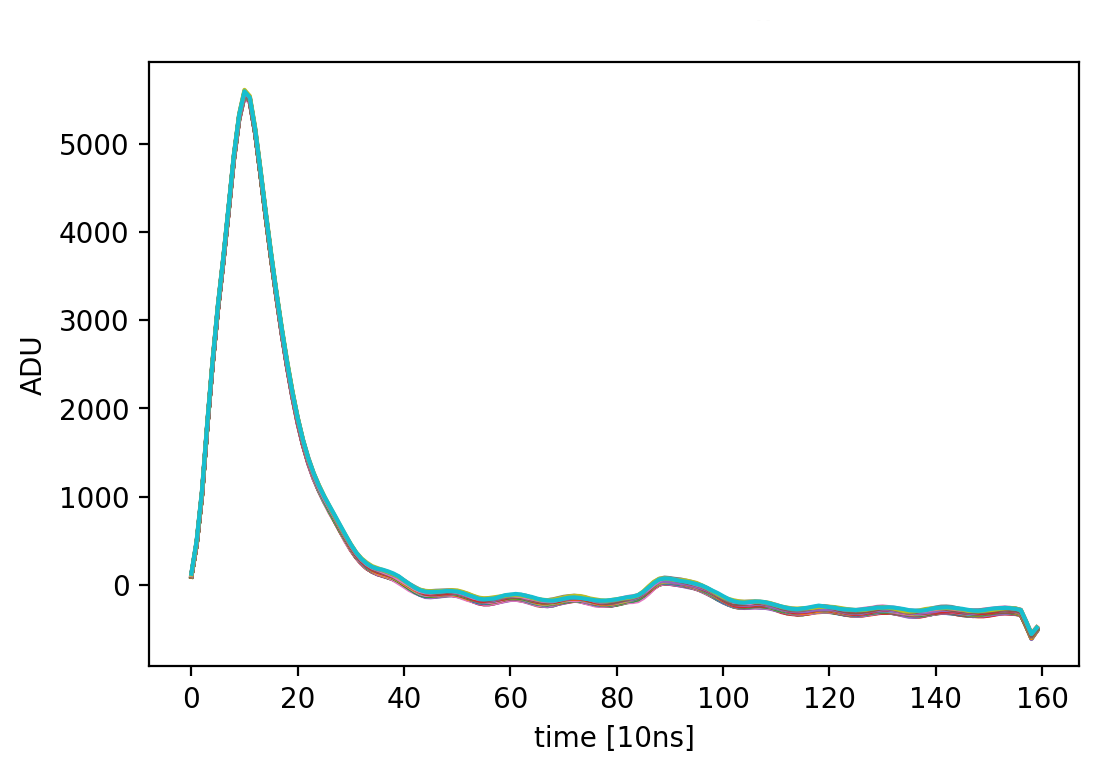}  
  \caption{ensemble of 50 waveforms (SiSeRO)}
  \label{fig:SiSeRO_WF_Raw}
\end{subfigure}
\begin{subfigure}{.51\textwidth}
  \centering
  \includegraphics[width=1.0\linewidth]{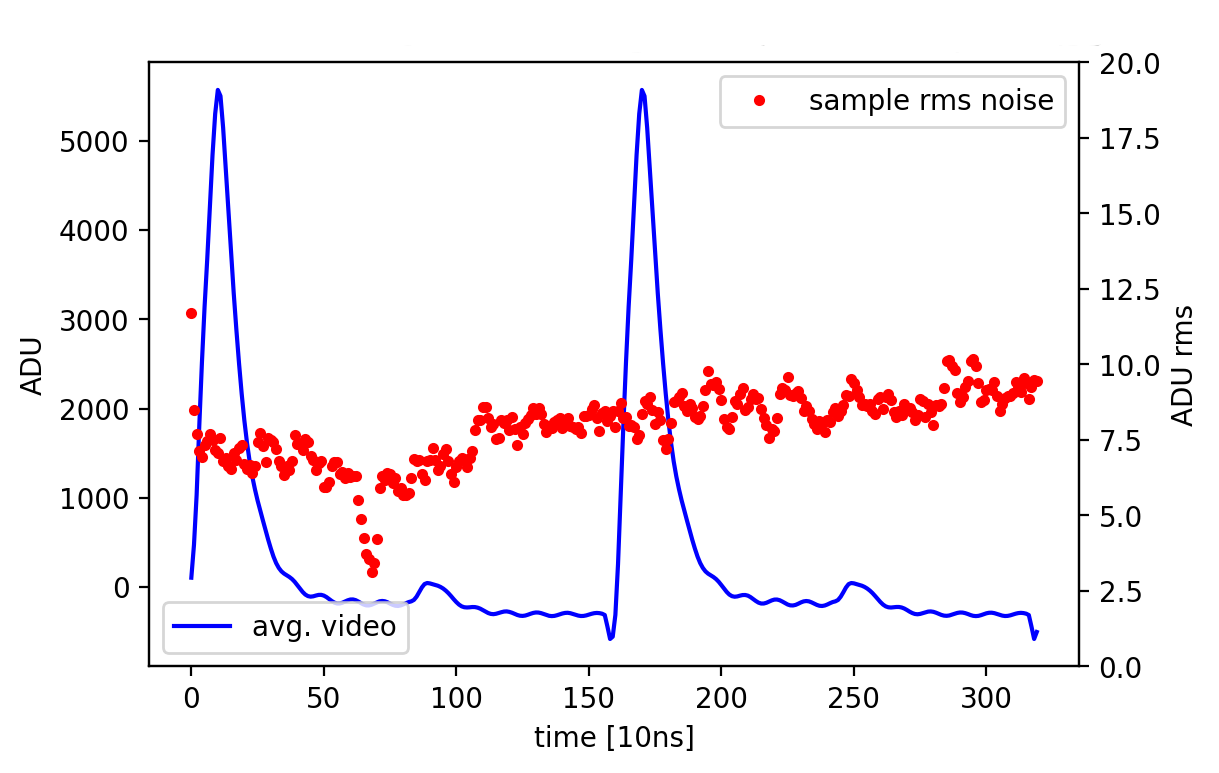}
  \caption{rms sample variation}
  \label{fig:SiSeRO_WF_RMS}
\end{subfigure}

\caption{
Left - Video waveform ensemble of 50 pixels with the SiSeRo output. The large pulse at the beginning is from the clear pulse, while the small bump around 90 ns is from the charge transfer into the back gate of the SiSeRO. The slight variation between the 50 waveforms can be seen. 
Right - Root-mean-square variation of the sample points across the 50 waveforms after each waveform has been clamped at the baseline level. 
Two successive waveforms are shown to illustrate the evolution across pixels.  Variation is low at the baseline level where the video waveform is clamped and increases left and right from this point. The charge clear pulse has no effect on the sample variation and no KTC noise bump is visible.}
\label{fig:SiSeRO_WF}
\end{figure}

\begin{figure} [ht]
\includegraphics[width=0.7\textwidth,center,keepaspectratio=tru]{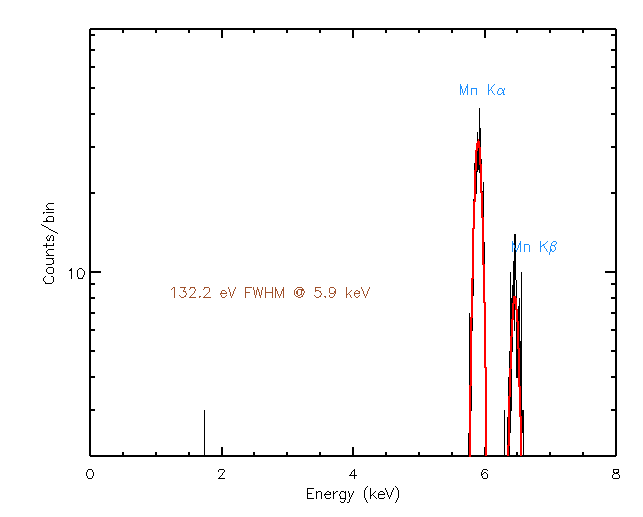}
\caption[SiSeROspectrum] 
{ \label{fig:SiSeROspectrum} 
The X-ray spectrum of a $^{55}$Fe radioactive source for single-pixel (grade 0) events measured at 625kPix/sec. As the SiSeRo device is KTC noise free, the measurement uses the baseline regions before and after the signal to estimate the baseline for improved noise performance (5 $e^{-}_{RMS}$).}
\end{figure}

Previously, we have demonstrated proof of principle measurements of novel SiSeRo devices fabricated by MIT Lincoln Laboratory \cite{chattopadhyay22_sisero}. In our ongoing characterisation efforts, we have used the same waveform analysis techniques outlined above to study the overall variation and reset behavior of these devices. As SiSeRO devices clear charge from the back gate region by simply transferring out, like any other charge transfer in a charge coupled device, this process should not exhibit KTC noise. Figure \ref{fig:SiSeRO_WF_Raw} shows an ensemble of 50 video waveforms from 50 different pixels (as above for the JFET in \ref{fig:JFET_WF_Raw}). Figure \ref{fig:SiSeRO_WF_RMS} plots the sample point variation of the waveforms after they have been referenced to the average value at the baseline. The plot shows two successive pixels to illustrate the evolution across pixels. As expected, no KTC noise bump is visible; instead, the variation increases slightly but steadily as the time separation to the reference level increases. This increase can be related to the substantial 1/f noise present in these early prototype MOSFET devices. Due to the SiSeRO device features, the signal measurement can use not only the baseline level before the signal but also the baseline level after the signal, or both. This technique of using both baseline levels increases the effective integration time for the digital CDS baseline and reduces the noise without a decrease in pixel rate: in our measurement, we achieved 6.2 $e^{-}_{RMS}$ noise for a classical digital CDS at 625 kPix/sec rate, while the noise improved to 5.0 $e^{-}_{RMS}$ by using the baseline before and after the signal (at the same 625 kPix/sec rate). Figure \ref{fig:SiSeROspectrum} shows the resulting X-ray spectrum from a Fe55 source. Based on this (almost) perfect transfer of the signal charge into and out of the back gate of the transistor we investigated repetitive non destructive readout (RNDR) techniques which can further improve the noise performance (at the expense of increased readout time): With nine repetitive readout cycles we achieve an noise as low as 2 $e^{-}_{RMS}$. Details and further results can be found in \cite{Chattopadhyayetal2022}. Once this technique is refined and reaches sub-electron noise levels, it could be of considerable advantage for the (in-situ) calibration of X-ray detectors \cite{2021NIMPA101065511R}.

\section{CONCLUSION}
Substantial progress has been made by our Stanford and MIT team to increase the frame rate of state-of-the-art X-ray CCD technology. We follow a two-pronged approach, working to increase both the readout speed of individual outputs and the number of outputs per CCD. We have demonstrated operation of MIT-LL CCID85 JFET output stages at 4 MPIX/sec per channel using discrete electronics, maintaining a spectral resolution of better than 150 eV FWHM for the 5.9 keV line. Improved output stages exhibit twice the conversion gain of the CCID85 and correspondingly improved noise performance:  2.6 $e^{-}_{RMS}$ at 2.0 MPix/sec \cite{Prigozhinetal2022}. To support multi-output CCDs, we have developed a dedicated readout ASIC, the MCRC V1 ASIC, which has 8 readout channels capable of operating at high speeds with modest power consumption. Full functionality of the MCRC V1 ASIC has been demonstrated \cite{Oreletal2022}, and tests of the paired 8-channel MIT-LL CCID89 + MCRC V1 ASIC are scheduled for this fall \cite{Bautzetal2022}. In addition, we are investigating a novel detector technology manufactured at MIT Lincoln Laboratory, the Single electron Sensitive Read Out (SiSeRO) that, while not yet (quite) capable of single electron noise performance, offers a promising pathway to very low noise,  high speed X-ray detectors. We have demonstrated that this output stage is KTC noise free and capable of repetitive nondestructive readout (RNDR) \cite{Chattopadhyayetal2022}.

\acknowledgments 
 
This work has been supported by the NASA APRA grant 80NSSC19K0499, “Development of Integrated Readout Electronics for Next Generation X-ray CCDs”, to Stanford  and the NASA SAT grant 80NSSC20K0401 “Toward Fast, Low-Noise, Radiation-Tolerant X-ray Imaging Arrays for Lynx: Raising Technology Readiness Further", to MIT.  


\end{document}